\documentstyle[aps,preprint]{revtex}
\tightenlines
\begin{document}

\draft

\title{An efficient projection algorithm and
its application to the spurious center-of-mass motion problem}

\author{H. Nakada$^1$, T. Sebe$^2$ and T. Otsuka$^3$}
\address{$^1$ Department of Physics, Chiba University,
Inage, Chiba 263, Japan\\
$^2$ Department of Applied Physics, Hosei University,
Koganei, Tokyo 184, Japan\\
$^3$ Department of Physics, University of Tokyo,
Hongo, Tokyo 113, Japan}

\date{\today}

\maketitle

\begin{abstract}
Based on the correspondence of projection
to an eigenvalue problem via the underlying group structure,
we generalize the projection algorithm proposed
by Morrison {\it et al}.
This algorithm is eligible for large-scale computations,
because we can avoid the accumulation of errors.
As an example, we apply it to the spurious c.m. motion
in a multi-$\hbar\omega$ shell-model calculation.
In comparison with Lawson's method,
the present method is exact and nevertheless
does not need longer CPU time.
It is pointed out that the exact treatment
of the spurious c.m. motion
is crucial to some physical quantities.
\end{abstract}

\pacs{PACS numbers: 02.70.-c, 21.60.Cs}

In the numerical treatment of a Hamiltonian in a large space,
we often need an efficient algorithm for projection,
in connection to conservation laws or spurious motions.
For instance, the $J$ (angular-momentum) projection accelerates
shell-model calculations in many cases.
A more serious problem is the spurious center-of-mass (c.m.) motion:
since we sometimes use a Hamiltonian
which may excite the spurious motion,
it is important to project it out of basis vectors.
In this report we extensively develop a projection algorithm,
which was first proposed in Ref.\cite{ref:MWW74}
for the $J$ projection,
and apply it to the c.m. motion problem
within the regime of multi-$\hbar\omega$ shell-model calculations.

Suppose that,
in an $N$-dimensional ($N$ is finite) Hilbert space $V$,
we have relevant and irrelevant components.
The projection operator $P$ picks up only the relevant components,
eliminating the irrelevant ones.
We now introduce a certain hermitian operator ${\cal H}_{\rm P}$,
which satisfies ${\cal H}_{\rm P}V\subseteq V$
and has degenerate eigenvalues within $V$.
We denote the eigenvalues by $\mu_0, \mu_1, \cdots, \mu_{n-1}$
($n<N$).
Eigenvectors associated with each eigenvalue $\mu_k$
span a subspace $W_k$ ($k=0,1,\cdots n-1$).
The full space $V$ is equal to the direct sum of $W$'s;
\begin{equation} V = \bigoplus_{k=0}^{n-1} W_k\;.
\label{eq:dir-sum}\end{equation}
Any vector $|v\rangle \in V$ is expanded as
\begin{equation} |v\rangle = \sum_{k=0}^{n-1} c_k
|v^{(k)}\rangle \;, \label{eq:expand}\end{equation}
where $|v^{(k)}\rangle \in W_k$
(i.e. ${\cal H}_{\rm P}|v^{(k)}\rangle = \mu_k |v^{(k)}\rangle$).
It is postulated that the relevant components belong
exclusively to $W_0$, which is attached to the eigenvalue $\mu_0$;
$PV=W_0$ and $P|v\rangle \propto |v^{(0)}\rangle$.
(We can straightforwardly extend our discussion to the case
that the relevant subspace is a sum of a few $W$'s.)
The projection can then be redefined as an eigenvalue problem
with respect to ${\cal H}_{\rm P}$, as will be illustrated below.

It is crucial to find
such an operator ${\cal H}_{\rm P}$ with a simple form.
The projection is normally linked to a certain group structure.
As is well-known, conservation corresponds to a symmetry,
and therefore to a group.
A spurious motion occurs when we ignore a symmetry to be satisfied.
This symmetry leaves its trace in a group structure
in the enlarged space.
In both cases, the relevant components are characterized
by a specific representation of the corresponding group.
Then a Casimir operator of the group
can be adopted as ${\cal H}_{\rm P}$;
for instance, ${\cal H}_{\rm P}=\hat{\bf J}^2$
in the $J$ projection.
The condition ${\cal H}_{\rm P}V\subseteq V$ is required
in order that the group structure holds within $V$.

For a given vector $|v\rangle \in V$,
$|v^{(0)}\rangle$ is obtained in the following manner,
by applying the Lanczos diagonalization algorithm\cite{ref:Wil65}
with respect to ${\cal H}_{\rm P}$.
Having a set of basis vectors
$\Gamma_{n'}\equiv \{ |v\rangle, {\cal H}_{\rm P}|v\rangle, \cdots,
({\cal H}_{\rm P})^{n'-1}|v\rangle \}$
with the Gram-Schmidt orthogonalization,
${\cal H}_{\rm P}$ is represented by a tridiagonal matrix.
We denote the $(i,j)$ component of this matrix
by $({\cal H}_{\rm P})_{ij}$.
At a certain $n'(\le n)$, $({\cal H}_{\rm P})^{n'}|v\rangle$ 
is no longer linearly independent of $\Gamma_{n'}$
because of the degenerate eigenvalues.
Then $|v^{(k)}\rangle$ is obtained
by solving a system of coupled linear equations
\begin{equation} \sum_j ({\cal H}_{\rm P})_{ij} x^{(k)}_j
= \mu_k x^{(k)}_i~(i=1,\cdots n')\;,
\label{eq:coupled} \end{equation}
where $x^{(k)}_i$ yields the expansion coefficient
of $|v^{(k)}\rangle$
for the Lanczos bases and $x_1^{(k)}=c_k$ of Eq.(\ref{eq:expand}).
Since ${\cal H}_{\rm P}$ is tridiagonal now,
the sum over $j$ on the left-handed side of Eq.(\ref{eq:coupled})
runs from $i-1$ to $i+1$ at most.
What we need is the eigenvector $|v^{(0)}\rangle$.
The probability $\langle v|P|v\rangle$ is given by $|x^{(0)}_1|^2$.
It has been known that solving Eq.(\ref{eq:coupled})
from $i=n'$ to $1$ with a small initial value of $x^{(0)}_{n'}$
is advantageous,
in order to keep rounding errors small\cite{ref:Wil65}.

Because of the group structure,
${\cal H}_{\rm P}$ has highly degenerate eigenvalues; $n\ll N$.
For the $J$ projection in shell-model calculations,
$n$ is the number of possible $J$ values in the space $V$,
being several tens even for the $M$-scheme shell-model space
with $N\sim 10^6$\cite{ref:NSO94}.
In the above algorithm,
we can keep the number of bases small ($n'\le n$).
As far as we know $\mu_0$ in advance,
the inverse iteration method is immediately applicable
to obtain $|v^{(0)}\rangle$.
However, solving Eq.(\ref{eq:coupled}) from $i=n'$ to $1$
is more efficient,
because we reach $|v^{(0)}\rangle$
with a small number of operations.
The Lanczos method has an advantage for these reasons,
although $n$ is not quite large.

It is noted that, by using ${\cal H}_{\rm P}$,
we can explicitly construct the projection operator $P$ as
\begin{equation} P = \prod_{k=1}^{n-1} \left(
{{{\cal H}_{\rm P} - \mu_k}\over{\mu_0 - \mu_k}}\right)\;.
\label{eq:P}\end{equation}
Each factor $({\cal H}_{\rm P} - \mu_k)/(\mu_0 - \mu_k)$
eliminates the irrelevant component $|v^{(k)}\rangle \in W_k$
with $k\ne 0$, leaving $|v^{(0)}\rangle$ at last.
Despite its mathematical exactitude,
this procedure is not numerically efficient.
First of all,
we always need an $n$-fold operation of ${\cal H}_{\rm P}$,
in contrast to $n' (\le n)$ in the present algorithm.
Moreover, in the process eliminating $|v^{(k)}\rangle$,
other irrelevant components may be enhanced,
causing serious numerical errors.
We take the $J=0$ projection as an example,
with ${\cal H}_{\rm P}=\hat{\bf J}^2$ and $\mu_k = k(k+1)$.
We operate $({\cal H}_{\rm P} - \mu_k)/(\mu_0 - \mu_k)$
on $|v\rangle$ from $k=1$ to $n-1$.
The irrelevant component $|v^{(1)}\rangle$
is removed at the first step.
However, if a small amount of the $W_{n-1}$ component
is contained in $|v\rangle$,
this component is enhanced by the factor $[n(n-1)-2]/2\sim O(n^2)$.
After several steps, the $W_{n-1}$ component
is repeatedly enhanced
and eventually dominates over the relevant component,
giving rise to a loss of precision for the relevant component
in practical floating-point computations.
It is numerically advantageous to enhance the relevant component,
which corresponds to a single eigenvalue $\mu_0$,
rather than to remove many irrelevant components.
In the present algorithm, the relevant component is enhanced
because of the order of solving Eq.(\ref{eq:coupled}).

Morrison {\it et al.} have pointed out the equivalence
between projection and an eigenvalue problem for some cases,
particularly for the $J$ projection\cite{ref:MWW74}.
They proposed the Lanczos algorithm
with ${\cal H}_{\rm P}=\hat{\bf J}^2$.
However, it has not been discussed how to obtain ${\cal H}_{\rm P}$
in general cases,
since one has not fully realized the underlying group structure
in the correspondence of the projection to an eigenvalue problem.
Furthermore, only the case of $[H,{\cal H}_{\rm P}]=0$
was discussed in Ref.\cite{ref:MWW74}.
Generalizing their argument,
we have now developed the projection algorithm systematically.

Since we apply the Lanczos method,
${\cal H}_{\rm P}$ as well as $H$ do not have to be represented
in a matrix form\cite{ref:WWCM}.
The present algorithm may be extended
even to the case that $n$ (and therefore $N$) is infinite
or the eigenvalues of ${\cal H}_{\rm P}$ are continuous.
In such cases, however, the degree of convergence
with finite $n'$ strongly depends on the vector $|v\rangle$.
We have restricted ourselves to the finite $N$ cases
in this article,
by which the convergence for $n'$ is assured.

We next demonstrate the present projection method
for the spurious c.m. motion
in a multi-$\hbar\omega$ shell-model calculation.
Shell-model calculations in multi-$\hbar\omega$ spaces
are under current interest, in relation to
a microscopic description of nuclei\cite{ref:micro}
and to a research on light unstable nuclei\cite{ref:unstab}.
Because the origin of the coordinate space
is fixed in the shell model, the c.m. motion,
which is irrelevant to the nuclear structure itself,
is included in the model space.
With the harmonic-oscillator (h.o.) single-particle wavefunctions,
the c.m. motion is also expanded
in terms of the h.o. representation, which forms a $U(3)$ group.
It is desired to constrain the c.m. mode to the $0s$ orbit.
For this purpose, we take
\begin{equation} {\cal H}_{\rm P}
= {1\over{2AM}}{\bf P}^2 + {{AM\omega^2}\over 2}{\bf R}^2
 - {3\over 2}\hbar\omega\;, \end{equation}
where ${\bf R}$ and ${\bf P}$ denotes c.m. coordinate and momentum.
This is indeed proportional to a Casimir operator
(linear Casimir operator) of the $U(3)$ group,
and is expressed with up to two-body operators
within the shell-model space,
\begin{equation} {\cal H}_{\rm P} = {1\over A}
\left[\sum_i\left({1\over{2M}}{\bf p}_i^2
+ {{M\omega^2}\over 2}{\bf r}_i^2 - {3\over 2}\hbar\omega\right)
+ \sum_{i<j}\left({1\over M} {\bf p}_i\cdot{\bf p}_j
+ M\omega^2 {\bf r}_i\cdot{\bf r}_j\right)\right]
\;. \end{equation}
The $0s$ mode corresponds to the eigenvalue $\mu_0=0$.

In order for the c.m. motion to be well-defined
inside the model space $V$,
${\cal H}_{\rm P}V\subseteq V$ should be satisfied.
Otherwise the model space should be extended.
We consider a model space with the $\hbar\omega$-type truncation;
$V$ is a direct sum of subspaces characterized
by the number of oscillator quanta $m$, $V=\bigoplus_m V^{(m)}$.
Each subspace $V^{(m)}$ is expanded
by direct products of the relative motion and the c.m. motion,
both of which have specific numbers of oscillator quanta,
$V^{(m)}=\bigoplus_k
[{\cal V}_{m-k}^{\rm rel}\otimes{\cal W}_k^{\rm c.m.}]$.
A resummation as $W_k=\bigoplus_\ell
[{\cal V}_\ell^{\rm rel}\otimes{\cal W}_{k}^{\rm c.m.}]$
leads to the expression of Eq.(\ref{eq:dir-sum}).
The space $W_k$ is characterized
by the symmetric representation $[k]$
of the c.m. $U(3)$ group
and has $\mu_k = k\hbar\omega$.
Obviously ${\cal H}_{\rm P}V\subseteq V$ is satisfied.
All the components outside $W_0$
are regarded as spurious components.
The corresponding projection operator $P$ of Eq.(\ref{eq:P})
has been discussed in Ref.\cite{ref:SKH}.
Unlike the $J$ projection,
we do not necessarily demand that the shell-model Hamiltonian $H$
should commute with ${\cal H}_{\rm P}$\cite{ref:GL74},
under the presence of a core or inactive particles.

\begin{figure}
\centering
\begin{picture}(205,205)
\put(1,175){Diag. of $PHP$}
\put(0,0){\dashbox{5}(205,172)}
\put(132,195){Diag. of ${\cal H}_{\rm P}$}
\put(102,144){\dashbox{5}(80,48)}
\put(123,182){$|v_1\rangle$}
\put(129,177){\vector(0,-1){12}}
\put(109,150){$|v_1^{(0)}\rangle\propto P|v_1\rangle$}
\put(47,153){\line(1,0){60}}
\put(18,150){$|v_1^{(0)}\rangle$}
\put(27,144){\line(0,-1){7}}
\multiput(26,131)(0,-3){3}{$\cdot$}
\put(27,124){\vector(0,-1){12}}
\put(18,100){$|v_i^{(0)}\rangle$}
\put(27,94){\line(0,-1){7}}
\put(27,87){\vector(1,0){76}}
\put(132,99){Diag. of ${\cal H}_{\rm P}$}
\put(99,46){\dashbox{5}(92,50)}
\put(106,84){$|\tilde v_{i+1}\rangle\equiv H|v_i^{(0)}\rangle$}
\put(131,79){\vector(0,-1){12}}
\put(106,52){$|\tilde v_{i+1}^{(0)}\rangle\propto P|\tilde v_{i+1}
\rangle$}
\put(27,55){\line(1,0){76}}
\put(27,55){\vector(0,-1){10}}
\put(15,33){$|v_{i+1}^{(0)}\rangle$}
\put(27,27){\vector(0,-1){12}}
\multiput(26,9)(0,-3){3}{$\cdot$}
\end{picture}
\vspace{5mm}
\caption{\label{fig:Lanc}
Illustration of the present projection algorithm
incorporated into the Lanczos diagonalization of $H$.
The basis vectors ($|v_i^{(0)}\rangle$) are generated
so as to stay inside $W_0$.}
\end{figure}
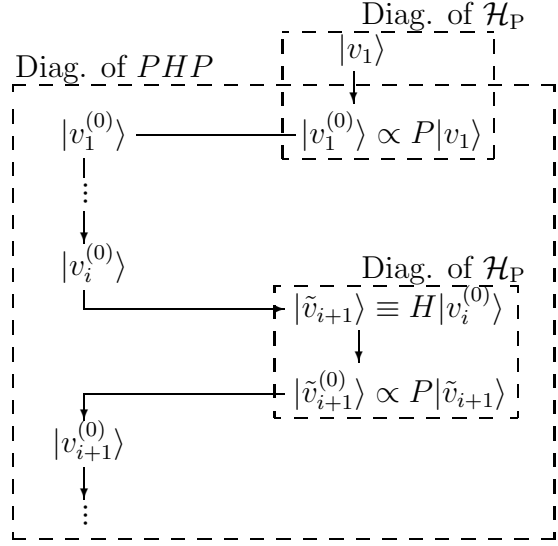

The present projection is incorporated
into the shell-model diagonalization.
The Lanczos method is usually employed
in diagonalizing a shell-model Hamiltonian\cite{ref:WWCM,ref:Lanc}.
The diagonalization of $H$ within $W_0$
is equivalent to the diagonalization of $PHP$ in $V$,
except that the irrelevant states have null eigenvalues for $PHP$.
Note that the explicit form of $PHP$ is complicated.
Given an initial vector $|v_1\rangle$,
we apply the present projection and obtain $|v_1^{(0)}\rangle$.
The next Lanczos basis $|v_2^{(0)}\rangle$ is
generated from $PH|v_1^{(0)}\rangle$.
Multiplication by $H$ is performed
just as in the usual shell-model calculation.
The projection on $|\tilde{v}_2\rangle \equiv H|v_1^{(0)}\rangle$
yields $|\tilde{v}_2^{(0)}\rangle$,
from which $|v_2^{(0)}\rangle$ is obtained
by orthogonalizing to $|v_1^{(0)}\rangle$.
The Lanczos bases of $PHP$ are generated one after another
in this manner,
as is summarized in Fig.~\ref{fig:Lanc}.
The computation proceeds according to the arrows.
Each Lanczos diagonalization is marked on the boxes.
At left the main process of the diagonalization of $PHP$
is indicated,
while the projection by diagonalizing ${\cal H}_{\rm P}$
is shown at right.

When the single-particle energies consist
only of kinetic energy and the effective interaction
depends only on relative coordinates with the h.o. bases,
the shell-model Hamiltonian can be made commutable
with ${\cal H}_{\rm P}$\cite{ref:Pal67}.
Even in such cases with $[H,{\cal H}_{\rm P}]=0$ (i.e. $[H,P]=0$),
the present projection is useful to cut off numerical errors.
It may be applied in the Lanczos iterations of $H$
at a certain interval, as well as to the initial vector.

For the spurious c.m. motion problem, the so-called Lawson's method
has conventionally been used\cite{ref:WWCM,ref:GL74,ref:Pal67}.
The original Hamiltonian is amended as
\begin{equation} H' = H + \lambda {\cal H}_{\rm P}
\quad(\lambda>0)\;.
\end{equation}
The ${\cal H}_{\rm P}$ term shifts up the spurious components
and reduces their admixture in low-lying states.
The quality of this prescription for large $\lambda$
can be assessed by perturbation theory.
By regarding $\lambda {\cal H}_{\rm P}$
as an unperturbed Hamiltonian
and $H$ as a perturbation,
an eigenvector of $H'$ is approximated by
\begin{equation} |v_\lambda\rangle \cong |v^{(0)}\rangle
- \sum_{k=1}^{n-1} \eta_k |v^{(k)}\rangle \;;~~
\eta_k = {1\over{\lambda k\hbar\omega}}
 \langle v^{(0)}|H|v^{(k)}\rangle
\sim O(\lambda^{-1}) \;.
\end{equation}
As far as $[H,{\cal H}_{\rm P}]\ne 0$,
$\eta_k$ does not vanish.
Notice that $\eta_1$ is the largest in general,
because of the energy denominator.
The corresponding energy is
\begin{equation} E_\lambda \equiv
\langle v_\lambda|H'|v_\lambda\rangle
\cong E^{(0)} - \sum_{k=1}^{n-1} \eta_k
\langle v^{(0)}|H|v^{(k)}\rangle + O(\lambda^{-2}) \;,
\end{equation}
where $E^{(0)} \equiv \langle v^{(0)}|H|v^{(0)}\rangle$.
Thus the influence of the spurious modes
on the wavefunctions and energies is proportional to $\lambda^{-1}$,
not diminishing rapidly as $\lambda$ increases.
On the other hand, as we set larger $\lambda$,
the convergence with respect to $H'$ becomes slower,
because the ${\cal H}_{\rm P}$ term dominates
and fine splitting due to $H$ is not easily attained.
Moreover, even if the influence of the spurious motion
is small for energies,
it is not necessarily negligible for transition properties,
particularly when there is some cancellation.

We have carried out a numerical test for the $^{10}$Be nucleus.
In the lowest configuration we have six nucleons in the $p$-shell,
on top of the $^4$He core.
We take the model space $V$ spanned by
up to the $3\hbar\omega$ excitations ($n=4$).
For $H$, the single-particle energies
are determined from the Woods-Saxon potential
with the parameters of Ref.\cite{ref:BM1}.
The Millener-Kurath\cite{ref:MK} and USD\cite{ref:USD}
two-body interactions
are taken for the $(s\mbox{-}p\mbox{-}sd)$-shell,
while the M3Y $NN$ effective interaction\cite{ref:M3Y} is used
for the rest.
This Hamiltonian is adopted just for a numerical test,
putting aside its physical appropriateness.
As a matter of fact,
the negative-parity levels have too high excitation energies.

As in Fig.~\ref{fig:Lanc},
the subroutine for the c.m. projection is incorporated
into the shell model code VECSSE\cite{ref:NSO94,VECSSE}.
The present projection technique is compared with Lawson's method.
For the $\lambda$ parameter in Lawson's method, we take 
0 (i.e. no care of c.m. motion), 1, 5 and 20,
with $\hbar\omega=45A^{-1/3}-25A^{-2/3}=15.5$~MeV.
The CPU time on HITAC S3800/480
is presented in Table~\ref{tab:CPU}.
Both the positive- and negative-parity states are computed
up to $J=4$, and 5--7 lowest eigenstates are computed
for each $J^P$,
with the convergence criterion of 0.1~keV.
While the present method gives an exact projection,
it does not make the CPU time so long.
In Lawson's prescription,
it takes a longer CPU time as $\lambda$ increases,
because more iterations (i.e. more Lanczos bases)
are necessary to fulfill the convergence criterion.
While for $\lambda=1$ the iteration number
is almost the same as that in the present projection,
it costs almost twice as many iterations for $\lambda=20$,
resulting in longer CPU time than the exact projection.
The calculated energies are presented in Table~\ref{tab:energy}.
The energy eigenvalues of Lawson's method do not rapidly converge
for increasing $\lambda$;
the 100~keV discrepancy remains even with $\lambda=20$.
The excitation energies, on the other hand, are rather stable.
For $\lambda=0$, the spurious component dominates
the lowest $1^-$ state with $E_x=13.998$~MeV,
while the third lowest $1^-$ state consists
mainly of the $W_0$ component,
whose energy is presented in Table~\ref{tab:energy}.
As a measure of admixture of the spurious components,
the expectation value of ${\cal H}_{\rm P}$ is
also shown for $4^+_1$.

We turn to the transition properties.
Some of the calculated results
are presented in Table~\ref{tab:trans}.
The $E1$ transition strengths are rather susceptible
to the contamination of the spurious c.m. motion,
since the bare $E1$ operator
$T(E1) = e\sqrt{3/4\pi} \sum_{i\in p} {\bf r}_i$
has the isoscalar part proportional to ${\bf R}$.
Here ${\bf r}_i$ denotes the coordinate of the $i$-th nucleon
and the sum runs over all the protons.
A simple modification of
$T'(E1) = e\sqrt{3/4\pi} \sum_{i\in p} ({\bf r}_i - {\bf R})$
reduces the influence of the spurious motion\cite{ref:EG2}.
However, we do not find apparent improvement
for Lawson's method in Table~\ref{tab:trans}.
On the other hand, the present method yields equal values
between $T$ and $T'$,
since the c.m. excitation is entirely removed.

A similar treatment for the $E2$ transition to $T'(E1)$
requires two-body operators, and hence is not popular.
We here assume the usual one-body $E2$ operator
with the bare charges.
Compared with the exact values,
Lawson's method works well in most cases.
However, the $1^-_2\rightarrow 2^-_4$ transition,
whose strength is not large, shows a significant discrepancy.
Even when excitation energies seem good for a physical discussion,
the influence of the spurious motion may be serious
for transition strengths, particularly for relatively small ones.

In summary, we have developed a projection algorithm,
extending the method of Ref.\cite{ref:MWW74}.
The projection is redefined as an eigenvalue problem
of a Casimir operator of the underlying group.
On this ground, the Lanczos diagonalization technique
can be utilized also for projection.
This algorithm is efficient
particularly in large-scale computations,
because numerical errors do not accumulate severely.
We have applied it to eliminate the spurious c.m. motion
in a multi-$\hbar\omega$ shell-model calculation.
Whereas the present method is exact,
it does not need longer CPU time than Lawson's method.
The present algorithm is useful, or even crucial,
to evaluate physical quantities
sensitive to admixture of the c.m. excitation.

$\\$
We thank A. Gelberg for careful reading the manuscript.
The Millener-Kurath and USD interactions
registered in the OXBASH program package\cite{OXBASH}
have been used, with the permission of B. A. Brown.

\begin{table}
\centering
\caption{CPU time (sec) for diagonalization of $H$,
with the c.m. motion handled by the present algorithm
or Lawson's method.
\label{tab:CPU}}
\begin{tabular}{rrrrr}
   \multicolumn{1}{c}{Projection} 
   & \multicolumn{4}{c}{Lawson's method}\\
   & \multicolumn{1}{c}{$\lambda=0$} 
   & \multicolumn{1}{c}{$\lambda=1$} 
   & \multicolumn{1}{c}{$\lambda=5$} 
   & \multicolumn{1}{c}{$\lambda=20$} \\
\hline
   $586$ & $523$ & $535$ & $560$ & $629$\\
\end{tabular}
\end{table}

\begin{table}
\centering
\caption{Eigenenergy of the ground state
and excitation energies (MeV).
$\langle {\cal H}_{\rm P}\rangle$ value is also shown for $4^+_1$
in the parenthesis.
\label{tab:energy}}
\begin{tabular}{crrrrr}
   & \multicolumn{1}{c}{Exact}
   & \multicolumn{1}{c}{$\lambda=0$}
   & \multicolumn{1}{c}{$\lambda=1$} 
   & \multicolumn{1}{c}{$\lambda=5$}
   & \multicolumn{1}{c}{$\lambda=20$} \\
\hline
   $E(0^+_1)$ & $-375.845$ & $-376.691$ & $-376.453$ & $-376.142$ & 
   $-375.950$\\
   $E_x(2^+_1)$ & $4.246$ & $4.200$ & $4.212$ & $4.229$ & 
   $4.240$\\
   $E_x(4^+_1)$ & $17.922$ & $17.617$ & $17.707$ & $17.821$ & 
   $17.888$\\
   $(\langle {\cal H}_{\rm P}\rangle / \hbar\omega)$ &
   $(0.0)$ & {\scriptsize $(1.3\times 10^{-1})$} &
   {\scriptsize $(1.5\times 10^{-2})$} &
   {\scriptsize $(3.3\times 10^{-3})$} &
   {\scriptsize $(3.9\times 10^{-4})$}\\
   $E_x(1^-_1)$ & $19.266$ & $19.125$ & $19.166$ & $19.263$ & 
   $19.269$\\
   $E_x(3^-_1)$ & $19.409$ & $18.234$ & $19.476$ & $19.445$ & 
   $19.422$\\
\end{tabular}
\end{table}

\begin{table}
\centering
\caption{$B(E1)$ ($e^2{\rm fm}^2$) and $B(E2)$ ($e^2{\rm fm}^4$).
\label{tab:trans}}
\begin{tabular}{crrrrr}
   & \multicolumn{1}{c}{Exact}
   & \multicolumn{1}{c}{$\lambda=0$}
   & \multicolumn{1}{c}{$\lambda=1$} 
   & \multicolumn{1}{c}{$\lambda=5$}
   & \multicolumn{1}{c}{$\lambda=20$} \\
\hline
   \multicolumn{1}{r}{$B(E1;0^+_1\rightarrow 1^-_2)$ by $T$} &
   $0.901$ & $0.292$ & $0.953$ & $0.888$ & $0.898$\\
   \multicolumn{1}{r}{by $T'$} &
   $0.901$ & $0.498$ & $0.690$ & $0.878$ & $0.896$\\
   $B(E2;0^+_1\rightarrow 2^+_1)$ & $13.167$ & $12.905$ & $12.949$ & 
   $13.045$ & $13.121$\\
   $B(E2;1^-_1\rightarrow 3^-_1)$ & $2.059$ & $1.807$ & $2.047$ & 
   $2.059$ & $2.059$\\
   $B(E2;1^-_2\rightarrow 2^-_4)$ & $0.011$ & $0.413$ & $0.265$ &
   $0.002$ & $0.004$\\
\end{tabular}
\end{table}

\end{document}